\documentclass[twocolumn,floatfix,preprintnumbers,aps,prd,showpacs,footinbib,amsmath,amssymb,amsfonts]{revtex4}
\usepackage{graphicx}
\usepackage[dvips,usenames]{color}
\usepackage{dsfont}
\usepackage{hyperref}

\def\flamoi{}

\newcommand{\YC}{C}

\newcommand{\etal}{{\it et al.}}

\newcommand{\be}{\begin{equation}}
\newcommand{\ee}{\end{equation}}
\newcommand{\bea}{\begin{eqnarray}}
\newcommand{\eea}{\end{eqnarray}}
\newcommand{\gr}[1]{\mathbf{#1}}

\newcommand{\ii}{\mathrm{i}}

\newcommand{\N}[1]{N_{#1}}
\newcommand{\uline}[1]{\underline{#1}}
\def\st{\sigma_{\mathrm T}}

\def\dd{{\rm d}}

\definecolor{Myblue}{rgb}{0,0,0}
\definecolor{Myred}{rgb}{0,0,0}

\def\iB{{\color{Myred}i}}
\def\jB{{\color{Myred}j}}

\def\iT{{{\color{Myblue}i}}}
\def\jT{{{\color{Myblue}j}}}
\def\kT{{{\color{Myblue}k}}}
\def\lT{{{\color{Myblue}l}}}
\def\mT{{{\color{Myblue}m}}}

\newcommand{\eq}{\textrm{eq}}
\newcommand{\lss}{\textrm{LSS}}

\newcommand{\kPlmn}[3]{{}^{#3} {\uparrow \!\!\!K}^{#2}_{#1}}
\newcommand{\kMlmn}[3]{{}^{#3} {\downarrow \!\!\!K}^{#2}_{#1}}

\def\gt{f_{T}}
\def\gtb{f_{\bar T}}
\begin{document}

\title[The $y$-sky]{The $y$-sky: diffuse spectral distortions of the cosmic microwave background}
\author{Cyril Pitrou}
\email{cyril.pitrou@port.ac.uk}
\affiliation{Institute of Cosmology and Gravitation,
Dennis Sciama Building, Burnaby Road, Portsmouth, PO1 3FX (United Kingdom)}

\author{Francis Bernardeau}
 \email{francis.bernardeau@cea.fr}
 \affiliation{CEA, IPhT, 91191 Gif-sur-Yvette c{\'e}dex, (France)\\
              CNRS, URA-2306, 91191 Gif-sur-Yvette c{\'e}dex (France)}

\author{Jean-Philippe Uzan}
\email{uzan@iap.fr}
\affiliation{Institut d'Astrophysique
de Paris, UMR-7095 du CNRS, Universit\'e Paris-VI Pierre et Marie
Curie, 98bis bd Arago, F-75014 Paris (France),\\
Department of Mathematics and Applied Mathematics,\\
University of Cape Town, Rondebosch 7701, Cape Town (South Africa)}

\begin{abstract}
The non-linear evolution of the energy density of the radiation induces spectral distortions 
of the cosmic microwave background both at recombination and during the reionization era. 
This distortion has the same spectral signature as the one produced by the re-scattering of 
photons by non-relativistic hot electrons, the thermal Sunyaev-Zeldovich effect, whose amplitude is quantified by a Compton $y$ parameter. 
A diffuse $y$-sky is then expected to emerge from mode couplings in the non-linear evolution of the cosmological perturbations and to superimpose to the point source contributions of galaxy clusters.  The equations describing the evolution of the $y$ field and a hierarchy 
governing its angular multipoles are derived from the second order Boltzmann equation. These equations are
then integrated numerically to obtain the first predicted power spectrum of the diffuse $y$-sky. It is found to be a remarkable tracer of the reionization history of the Universe.
\end{abstract}

\maketitle
\section{Introduction}

The spectrum of the CMB temperature has been observed to be extremely close to
a Planck [black-body (BB)] spectrum \cite{1996ApJ...473..576F}. According to the standard lore,
the radiative transfer of the cosmic microwave background (CMB) does not induce any spectral distortions so
that the brightness of the radiation  propagating in given direction $n^\iT$ can be described by a BB spectrum.
It follows that fluctuations in the brightness map into temperature fluctuations. Spectral
distortions are mostly associated with secondary anisotropies and 
due to the interaction of CMB photons
with the electrons of the cosmic plasma with a higher temperature or due to
electron scattering on charged particle. They are respectively referred to as
thermal {\em Sunyaev-Zel'dovich} (SZ) and {\em free-free} and are both late time effects~\cite{SZoriginal,Vishniacs,Aghanim2007,houches}. Any energy injection before recombination and at redshifts smaller than $\simeq10^6$ would also reflect itself through a spectral distortion described by an effective chemical potential.

The previous statement is however stricly correct up to first 
order in cosmological perturbations. In general, fluctuations in the radiation distribution are
induced by gravitational effects or through Compton scattering out of free charged particles (mainly free electrons). 
While gravitational effects do not introduce spectral distortions
at any order in perturbations -- since the geodesics of photons with proportional 4-momenta are 
actually the same -- Compton scattering does generate distortions beyond the first order
in perturbation theory. The spectral
distortion arises mostly at the last scattering surface (LSS) and during reionization, 
from the electron flows {\em irrespectively} of their thermal velocity dispersion.
We are thus dealing with a  non-linear kinetic SZ effect, physically different from a thermal SZ effect. However, as we shall see, it has the same spectral signature.

The emergence of a diffuse component in the spectral distortion is thus a generic feature of 
the cosmological dynamics beyond the linear order. It has so far been ignored in second order 
effects treatments.
The goal of this article is to show explicitly that this spectral distortion is 
characterized by a Compton $y$ parameter, that is similar to
the termal SZ effect, and then to derive its evolution equation from
the general second-order Boltzmann equation~\cite{Pitrou2008} (from which we use the results and notation). 
We then estimate its amplitude and distribution (through its angular power spectrum). As we whall discuss, this offers
potentially a new window on the dynamics of the cosmological perturbations and on the reionization era.

\section{Describing spectral distortion}

Let us first describe the spectral disortion.
In the Fokker-Planck approximation, that is at lowest order in the distortions, 
the distribution function of radiation $I(E)$ around a background BB spectrum can be expanded 
as~\cite{Stebbins2007,Challinor1997}
\be\label{EqDefy1}
I(E)=\gt(E)+ y {E^{-3}}D\left[ E^3 D \gt(E) \right]\,,
\ee
with $D\equiv E \frac{\partial}{\partial E}$, and where $\gt(E)$ is the BB spectrum at temperature $T$ which depends only on $E/T$. 

This decomposition defines the temperature  $T=\bar T (1+\Theta)$ which is the temperature of the BB which would have the same number density of photons, and we can name it the {\em occupation number} temperature. However, it is important to realize that other definitions of the temperature are possible. For instance, one can define a temperature out of the brightness ${\cal I}\propto \int I(E)E^3 \dd E$. On the background spacetime, the value of the brightness is directly related to the energy density of radiation since $\bar{\cal I}=\bar\rho_\gamma$, and we can define the brightness temperature as the temperature of the BB which would have the same brightness as the actual distribution. One would thus define the {\em bolometric} temperature $T_B$ by~\cite{PUB2}
\be\label{defTbolometrique}
\left(\frac{T_B}{\bar T}\right)^4\equiv \frac{{\cal I}}{\bar{\cal I}}\,\,.
\ee 
Indeed both at the background and at first-order level, both definitions reduce to the same quantity since in each direction of propagation the spectrum is still the one of a BB. However this is not the case as soon as collisions induce spectral distortions, that is at second-order in perturbations and beyond. According to the decomposition~(\ref{EqDefy1}), these two definitions are related through
\begin{equation}\label{defToccupation}
 T\equiv \frac{T_B}{(1+4 y)^{\frac{1}{4}}}\,\,.
\end{equation}
The decomposition~(\ref{EqDefy1}) defines the $y$-type distortion which is the only distortion generated by second order cosmological peturbations. 
Such a distortion modifies the energy density of the radiation but not the photon number density, contrary to the photons produced
during recombination which conserve none of these densities~\cite{peebles}.
In general the distribution of radiation depends on the coordinates $(\eta,x^\iB)$ in space-time and on the 
direction of propagation $n^\iT$. The temperature of the BB spectrum is in principle different in each direction. 
We will often omit the $(\eta,x^\iB)$ dependence when no confusion can arise and shall express all BB spectra in terms 
of the background BB spectrum of temperature $\bar T(\eta)$. The temperature can be related $\bar T$ through the fractional temperature perturbation defined by
$T(\eta,x^{\iB},n^\jT)=\bar T(\eta) \left[1+\Theta(\eta,x^{\iB},n^\jT)\right]$. We can then relate the BB spectra through the 
Taylor expansion
\be
\gt(E)=\sum_{n=0}^{\infty}\frac{(-1)^n}{n!}[\ln(1+\Theta)]^n D^n \gtb(E)\,.
\ee
Up to second order in perturbations, we thus have
\bea\label{decI}
I&=&\gtb- \left(\Theta -\frac{1}{2}\Theta^2\right)D \gtb +\frac{1}{2} \Theta^2 D^2 \gtb  \nonumber\\
&+&y E^{-3}D[E^3 D\gtb]\,.
\eea
In this expansion, the position, time and direction dependence of $I(E)$ is to be found in $\Theta$ and $y$. We decompose the directional dependence of $\Theta$ in symmetric trace-free (STF) tensors 
(see Refs.~\cite{Pitrou2008,Pitrou2008grg} for details) as
$
\Theta(n^\iT)=\sum_{\ell=0}^\infty \Theta_{\uline{\iT_\ell}}n^{\uline{\iT_\ell}}
$,
where the $\Theta_{\uline{\iT_\ell}}\equiv \Theta_{\iT_1\dots\iT_\ell}$ are STF and where, conventionally, 
the lowest multipole (i.e. corresponding to $\ell=0$) is noted $\Theta_\emptyset$.
Since polarisation is also generated at first order, we develop the distribution tensor of 
linear polarisation $P_{\iT \jT}$ in electric and magnetic type multipoles 
\be\label{Def_multipoleP}
\flamoi P_{\iT \jT}({\bf n})=\sum_{\ell=2}^\infty \left[E_{\iT \jT
  \uline{\kT_{\ell-2}}}n^{ \uline{\kT_{\ell-2}}} \,- n_{\mT}\epsilon^{\mT
  \lT}_{\,\,\,\,\,(\iT}B_{\jT)\lT \uline{\kT_{\ell-2}}}n^{\uline{\kT_{\ell-2}}}\right]^{\mathrm{TT}}\,,\nonumber
\ee
where ${\mathrm{TT}}$ means that we extract the trace-free part transverse with respect to $n^\iT$. 
In practice, we only need the polarisation at first order and, since the magnetic 
multipoles are not generated in the absence of primordial vector and tensor modes, 
we can safely neglect them.

\section{Evolution of the spectral distortion}

The evolution of $y$ up to second order in perturbations can then be deduced directly from
the Boltzmann equation~\cite{Stebbins2007} in which we insert the expansion~(\ref{decI}). 
Including polarisation (see Ref.~\cite{Pitrou2008},
dropping the hatted notation to alleviate the equations below), it reduces to
\bea\label{Evoly}
\frac{\dd y}{\dd \eta}&=&\tau'\left[ -\tilde y+ \tilde y_\emptyset+\frac{1}{10}\tilde y_{\iT \jT}n^\iT n^\jT + (\Theta- v_\iT n^\iT) (\Theta-\Theta_\emptyset) \right. \nonumber\\
&&\quad -\frac{1}{10}\Theta_{\iT \jT}n^\iT n^\jT \Theta-\frac{3}{10}\Theta_\iT v^\iT-\frac{1}{10}\Theta_\iT v_\jT n^\iT n^\jT\nonumber\\
&&\quad + \frac{1}{10}\Theta_{\iT \jT} v_\kT n^\iT n^\jT n^\kT -\frac{3}{70}\Theta_{\iT \jT \kT} n^\iT n^\jT v^\kT\nonumber\\
&&\quad+\frac{1}{3}v_\iT v^\iT +\frac{11}{20} v_\iT v_\jT n^{\langle \iT} n^{\jT \rangle}\\
&&\quad\left.-\frac{3}{5}E_{\iT \jT} (v_\kT n^\kT-\Theta)n^\iT n^\jT+\frac{1}{7}E_{\iT \jT \kT} n^\iT n^\jT (v^\kT-\Theta^\kT) \right]\,,\nonumber
\eea
where $\tilde y\equiv y+\Theta^2/2$; $\langle \iT_1 \dots \iT_n\rangle$ denotes 
the symmetric trace-free part of a tensor; $\tau'$ is the interaction rate, given by 
$\tau'=a \bar n_{\rm e} \st$  where $\bar n_{\rm e}$ is the background density of 
free electrons and $\st$ the Thomson scattering cross section; $v^\iT$ is the velocity of baryons in 
the Poisson (or longitudinal) gauge~\cite{Pitrou2008}. Note that only $y$ is a second order quantity so that all other quantities, appearing in quadratic terms, have to be taken at first order only. We thus omit the orders of perturbations as no confusion can arise. We emphasize that in the r.h.s. $\tau'$ multiplies a second order quantity so that it has to be evaluated
at the background level, and the fluctuations of $n_e$, responsible e.g. or the
Ostriker-Vishniac effect~\cite{Aghanim2007}, enter only at third order in perturbations. Furthermore, 
$y$ is a gauge invariant variable as pointed out in Ref.~\cite{Stebbins2007} and as
can be checked directly from the results of Ref.~\cite{Pitrou2007} combined with the decomposition~(\ref{decI}). Consequently, both the right- and the left-hand sides of Eq.~(\ref{Evoly}) are independently gauge invariant. 
More precisely in this expression, the variables $\Theta$, $\Theta_{\uline{\iT_{\ell}}}$, $E_{\uline{\iT_{\ell}}}$ and $v^\iT$ are gauge invariant quantities which reduce to their usual perturbation counterparts in the Poisson gauge.
A further simplification arises since $y$ is a pure second order quantity and vanishes at the background and first order perturbation level:
$\dd/\dd\eta \equiv \partial/\partial \eta + (\partial x^\iB/\partial \eta) \partial /\partial x^\iB+ (\partial n^\iT/\partial \eta) \partial /\partial n^\iT$
needs only be evaluated at the background level, i.e.  as $\overline{\dd/\dd\eta} \equiv \partial/\partial \eta + n^\iT \partial /\partial x^\iB$.

We can extract the moments of the evolution equation~(\ref{Evoly}) to
obtain a hierarchy which would be solvable numerically. To simplify
the expressions obtained, we neglect the multipoles of $\Theta$ and $E$
for $\ell\ge 3$. We obtain
\bea
\frac{\partial y_{\uline{\iT_{\ell}}}}{\partial \eta}+\frac{\ell+1}{(2 \ell +3)}\partial^\jB y_{\jT
  \uline{\iT_\ell}} + \partial_{\langle \iB_\ell} y_{\uline{\iT_{\ell-1}}\rangle}=\tau' \YC_{\uline{\iT_{\ell}}}\,
\eea
where 
\bea
\YC_{\emptyset}
&=&\frac{1}{3}(\Theta_\iT-v_\iT)(\Theta^\iT-v^\iT)+\frac{3}{25}\Theta_{\iT\jT}\Theta^{\iT
\jT}\,,\\
\YC_{\iT} &=&-y_\iT+\frac{9}{25} \left(\Theta_{\iT \jT}+\frac{2}{3} E_{\iT \jT}\right)(\Theta^\jT-v^\jT) \,,\label{e8}\\
\YC_{\iT \jT}&=&-\frac{9}{10}y_{\iT \jT}+\frac{11}{20} \left(\Theta_{\langle \iT}-v_{\langle \iT}\right)\left(\Theta_{\jT \rangle}-v_{\jT \rangle}\right)+\frac{9}{35}\Theta_{\iT\kT}\Theta^{\kT\jT}\nonumber\\
\YC_{\iT \jT \kT}&=&-y_{\iT \jT \kT}+\frac{1}{10}\left(
  \Theta_{\langle \iT \jT} +\frac{2}{3} E_{\langle \iT \jT}\right)(\Theta_{\kT
  \rangle}-v_{\kT \rangle})\,,\\
\YC_{\iT \jT \kT\lT}&=&-y_{\iT \jT \kT\lT}+\frac{2}{5}\Theta_{\langle\iT\jT}\Theta_{\kT\lT\rangle}\,,\\
\YC_{\uline{\iT_{\ell}}}&=&-y_{\uline{\iT_{\ell}}} \qquad {\rm for} \qquad \ell \ge 5\,\,.
\eea
In the unpolarized case, this result is consistent with Eq.~(47) of Ref.~\cite{Stebbins2007}, once the integration over the scattering terms is performed. 

In the tight coupling approximation (that is at lowest order in powers of $k/\tau'$) the first order Boltzmann equation implies 
that  $\Theta_{\uline{\iT_{\ell}}} \propto \Theta_\iT (k/\tau')^{\ell-1}$ for $\ell \ge 2$. In this approximation, we can thus 
discard these quadrupolar terms and simplify acccordingly the collision multipoles.
The tight coupling approximation was shown to be accurate, at second order in perturbations,
during the recombination~\cite{pub1}. During reionization, the same approximation can be
performed, not because tight coupling holds, but mostly because the multipoles
$\Theta_\ell$,  $E_\ell$ and $B_\ell$ behave approximately like $j_\ell(k \eta)$
whereas $v^\iT$ grows since recombination and terms quadratic in the baryons velocity thus dominate. 
Again, the spectral distortion considered during reionization is thus a (non-linear) kinetic SZ effect.

\section{The distortion hierarchy}

For a numerical integration, it proves more convenient to use the basis of spherical harmonics rather than the STF tensors to develop the angular dependence. When working in Fourier space, we choose to align the azimuthal angle with the Fourier mode $\gr{k}$. 
Following the convention of Refs.~\cite{Pitrou2008,Hu1997}, we expand on the basis
\be\label{DefGlm}
G_{\ell m}(\gr{k},x^\iB,\gr{n})=(N_\ell)^{-1} e^{\ii
k_\iB x^\iB} Y^{\ell m}(\gr{n}) \,,
\ee
with $\N{\ell} \equiv \ii^\ell \sqrt{(2\ell+1)/(4 \pi)}$ and obtain our master equation
\be
\frac{\partial y_{\ell}^m(k)}{\partial \eta}+k\left[\frac{\kPlmn{\ell+1}{m}{0}}{2\ell+3}y_{\ell+1}^{m}(k)-\frac{\kMlmn{\ell}{m}{0}}{2\ell-1}y_{\ell-1}^{m}(k) \right]=\tau' \YC_\ell(k)\,,\nonumber
\ee
with the collision moments given by
\bea
\YC_0(k)&=&-y_0(k),\nonumber\\
  &&-{\cal K}\left\{\sum_{n=-1}^{1}\frac{(-1)^n}{3} V_n(\gr{k}_1)V_{-n}(\gr{k}_2),\right\}\\
\YC_1^m(k)&=&-y_1^m(k),\\
\YC_2^m(k)&=&-\frac{9}{10}y_2^m(k)\nonumber\\
&&+\frac{11}{20}{\cal K}\left\{\sum_{n=-1}^1
\frac{\kMlmn{2}{m}{n}}{3} V_{m-n}(\gr{k}_1) V_n(\gr{k}_2) \right\}\\
\YC_\ell^m(k)&=&-y_\ell^m(k)\,,
\eea
where we have defined the convolution by
\be
 \mathcal{K}\{\dots\} \equiv \int \frac{\dd^3\gr{k}_1
 \dd^3\gr{k}_2}{(2 \pi)^{3/2}}\, \delta_{\rm D}^{3}(\gr{k}_1+\gr{k}_2-\gr{k})\dots\,
\ee
and with
\bea
V_m(\gr{k})&\equiv& v_m(\gr{k})-\Theta_1^m(\gr{k})\,,\\
\kPlmn{\ell}{m}{0}&\equiv& \kMlmn{\ell}{m}{0}\equiv
\sqrt{\frac{(\ell^2-m^2)}{\ell^2}}\,,\\
\kPlmn{\ell}{m}{\pm1}&\equiv& -\sqrt{\frac{(\ell \pm m)(\ell\pm  m
    +1)}{2 \ell^2}}\,,\eea\bea
\kMlmn{\ell}{m}{\pm1}&\equiv& \sqrt{\frac{(\ell\pm m)(\ell\pm m
    -1)}{2 \ell^2}}\,.
\eea

\section{The angular spectrum of distortions}

The spectral distortions are generated only when the
collisions become efficient enough, that is
when baryons and photons start to decouple and have different velocities. 
For adiabatic initial conditions, the velocities of all fluids are initially equal and suppressed on large scales, so that no significant 
spectral distortion is expected on large scales. 
We can thus safely compute the spectrum in the flat-sky approximation, 
even at the reionization time, given that it is reliable beyond $\ell\simeq10$ (see Ref.~\cite{PUBenpreparationflatsky}
for more details on this approximation and its domain of validity). 
The sources $S_\ell^m$, or more precisely the transfer function from the primordial potential $\Phi_{\rm i}$, on which we integrate along the line of sight are given by the implicit definition 
$
{\cal K}\left\{S_\ell^m(\gr{k}_1,\gr{k}_2)\Phi_{\rm i}(\gr{k}_1)\Phi_{\rm i}(\gr{k}_2)\right\} \equiv g(\eta)[\YC_\ell^m(k)+y_\ell^m(k)]
$
where  $\tau$ is the optical depth and satisfies $\partial \tau/\partial \eta=-\tau'$ with $\tau_0=0$, and 
$g(\eta)=\tau'e^{-\tau}$ is the visibility function. It receives two contributions: one which peaks in the middle of the LSS at a comoving distance which is by definition $r_{\rm LSS}$, and one which grows for $z\le 11$, due 
to the reionization of the universe. For the first contribution, the spectrum in the flat-sky approximation reads,
\bea
&&C_\ell^{yy} 
= \frac{1}{r^2_{\rm LSS}}\int \frac{\dd {k_2}_r \dd {k_1}_r  \dd {k_1}_\perp}{(2 \pi)^3} k_1 \sin \alpha_{\gr{k}_1,\gr{k}} P(k_1)P(k_2)\nonumber\\
&&\times  \sum_m \left|\int \dd r g(r) \sum_{\ell} Q_\ell^m(k_r/k) S_\ell^m(\gr{k}_1,\gr{k}_2) e^{\ii k_r r}\right|^2\,.
\eea
In this expression, we have used the functions
\be
Q_\ell^m (k_r/k) \equiv \ii^\ell \sqrt{\frac{(\ell-m)!}{(\ell+m)!}}P_\ell^m\left(k_r/k\right) \,,
\ee
where $P(k)$ is the usual initial power spectrum of the primordial gravitational potential $\Phi_{\rm i}$ and $P_\ell^m(\cos \beta)$ are the associated Legendre polynomials, in agreement with the basis~(\ref{DefGlm}) used for the multipoles expansion. The Fourier modes $\gr{k}$ are split into a part $k_r$ along the line of sight and a part $k_\perp$ orthogonal to the line of sight, the latter satisfying the flat-sky constraint $k_\perp = (\ell+1/2) r_{\rm LSS}$~\cite{PUBenpreparationflatsky}.
It is also understood that $\gr{k}_2=\gr{k}-\gr{k}_1$, and $\alpha_{\gr{k}_1,\gr{k}}$ is the angle between $\gr{k}_1$ and $\gr{k}$. Finally, the distance $r$ of an emitting point is defined through $r(\eta) \equiv \eta_0-\eta$.

The second contribution arising from the reionization era can be evaluated using a Limber approximation which is a refinement of the flat-sky approximation valid when the sources are slowly varying and contribute in a wide range of distances, which is our case here. We obtain
\bea
&&C_\ell^{yy} 
= \frac{1}{(2 \pi)^2}\int \dd r \dd {k_1}_r \dd {k_1}_\perp  \left|{k_1}_r\right| P(k_1)P(k_2) \nonumber\\
&&\qquad \qquad  \times  \sum_m \left| \frac{g(r)}{r} \sum_{\ell} Q_\ell^m(0) S_\ell^m(\gr{k}_1,\gr{k}_2) \right|^2\,,
\eea
in which it is understood that $k_r=0$, that is ${k_2}_r=-{k_1}_r$ and $k_\perp =(\ell+1/2)/r$.

\begin{figure}[t]
\includegraphics[width=8cm]{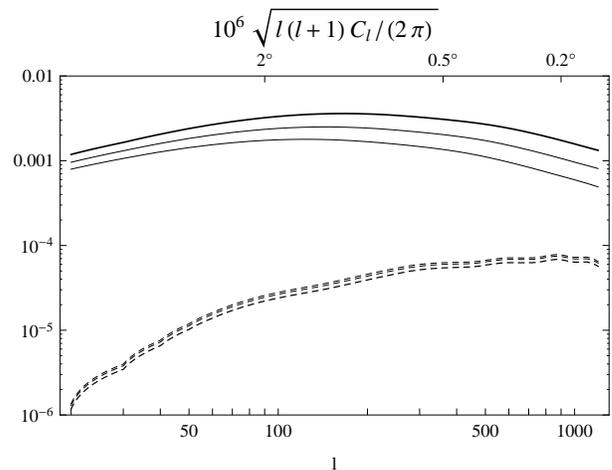}
\caption{The angular power spectrum for the $y$ distortion ($C^{yy}_\ell$) induced
at recombination (bottom dashed curves) and the signal induced at reionization (upper solid curves).
We report the contributions obtained for $\tau_{\rm reion}=0.043$, $0.087$ and $0.174$, drawing them
with increasing thickness.}
\label{figCl1}
\end{figure}

The results obtained are reported on Fig.~\ref{figCl1},
assuming the cosmological parameters of WMAP-5~\cite{WMAP5} with instantaneous and full reionization, 
but allowing the optical thickness to vary by a factor 2 compared to its WMAP-5 value. 
The distortion generated during reionization strongly dominates over the one arising from the LSS. 
Assuming that between the LSS and reionization the baryons are completely
decoupled, their velocity perturbation scales as $v^\iT \sim \Phi k\eta$ so that the
ratio between the signal from the reionization and the LSS is expected to be of order
$\tau_{\rm reion} (\eta_{\rm reion} / \eta_\lss)^4$. Given that
$\eta_{\rm reion}  \sim 50 /k_\eq$ and $\eta_\lss\sim 3 /k_\eq$, this harsh approximation~\footnote{Since, for a given 
$\ell$, the modes $k$ contributing to $C_\ell$ at different distances are not the same.}
leads to the rough estimates $C_\ell^{yy}({\rm reion})/C_\ell^{yy}(\lss)\sim{\cal O}(10^4)$ 
in full agreement with Fig.~\ref{figCl1}. Note also that $C_\ell^{yy}(\hbox{reion})$
does not scale linearly with $\tau_{\rm reion}$, mainly because increasing the
reionization means that it starts earlier but the higher-$r$ part of the integral
contributes less given that the integrand roughly scales as
$P^2 v^2/r^2\sim (\eta_0-r)^4$.\\

While $C_\ell^{TT}=C_\ell^{T_B T_B}$, the fact that $T$ and $T_B$ do not coincide at second order in perturbations implies that we can define two types of bispectra, $B^{T_{\rm B}T_{\rm B}T_{\rm B}}_{\ell_1 \ell_2 \ell_3}$  and $B^{TTT}_{\ell_1 \ell_2 \ell_3}$. At lowest order in spectral distortions, they are related by
\be
B^{TTT}_{\ell_1 \ell_2 \ell_3}=B^{T_{\rm B}T_{\rm B}T_{\rm B}}_{\ell_1 \ell_2 \ell_3}-B^{yTT}_{\ell_1 \ell_2 \ell_3}-B^{yTT}_{\ell_2 \ell_3 \ell_1}-B^{yTT}_{\ell_3 \ell_1 \ell_2}\,.
\ee
This indicates clearly the ambiguity of using one temperature rather than the other one because of spectral distortions. Fortunately, as we have just shown for the spectrum, the spectral distortions are mostly induced during the reionization epoch during which the terms of the form $B^{yTT}_{\ell_1 \ell_2 \ell_3}$ are suppressed by the fact that the main effect in the temperature is coming from the Doppler effect. Indeed, the Limber approximation tells us that most Fourier modes contributions come from the modes satisfying $k_r \ll k$ when the contributions are extended in a wide region in time, whereas the Doppler effect is more efficient when $k_r \sim k$, and this brings a geometric suppression of the coupling of spectral distortions to the Doppler effect during the reionized epoch. We thus expect the two bispectra to agree numerically, even though they do not mathematically.
\section{Conclusion} 

In this article, we have shown that, beyond the linear
order in cosmological perturbations, the fluctuations
in the CMB cannot be entirely described by a temperature
fluctuation since there is a generic y-type departure
from a BB spectrum ~\footnote{In general, at the $n$-th order in perturbation theory, the spectral dependence of each multipole of the CMB sky needs to be described as the superposition of $n$ different functions.}. We have derived the evolution
equation of this spectral distortion at second order
in perturbations from the second order Boltzmann equation
and we have emphasized that it leads to ambiguities
in the way the temperature is defined. As a consequence the actual temperature 
bispectrum resulting from the second order coupling depends 
on which temperature is considered (in practice though, we found the differences to be very weak).

The spectral distortion is found to be a
non-linear kinetic SZ which has the same spectral signature
as a thermal SZ effect (the linear kinetic SZ effect
inducing no disortion). We also remind that it is different
from a modulation of the Doppler effect from density
and ionisation variation~\cite{jubas}. The amplitude of this disortion
has been estimated numerically and we have shown that
the contribution from the reionization era dominates the
one arising from the recombination by almost 4 orders of
magnitude, a consequence of the late time acceleration
of the electrons. The properties of the y-sky are then
expected to depend mostly on the physics of reionization,
in a way that can turn to be much more precise
than the CMB polarization~\cite{2003ApJ...595...13H}. However, before drawing
definitive statements, it is necessary to show that such
effects can be distinguished from the thermal SZ due to
hot electrons in galaxy cluster halos. This latter effect
is expected to produce an average signal larger than the
signal described here. Although tentative to disentangle the CMB contribution from the
galactic emission and the unresolved extragalactic radio
sources contributions of the sky temperature has been made~\cite{tris2}, it is clear that
whether those point sources could
in the future be identified in a large enough number to
separate the two contributions is still an open issue.

{\it Acknowledgements:} C.P. is supported by STFC. F.B. thanks ICG and C.P. thanks IAP for their kind hospitality during the preparation of this work.


\end{document}